\documentclass[11pt]{article}
\usepackage{amsmath,amssymb,amsthm,empheq}
\usepackage[dvipdfmx]{graphicx}
\usepackage{comment}
\usepackage{fancybox}
\usepackage{eurosym}
\usepackage{type1cm}
\usepackage{color}
\newtheorem{theorem}{Theorem}[section]

\newtheorem{lemma}[theorem]{Lemma}

\usepackage{bm}

\newcommand{\QED}{\hfill $\Box$}
\newcommand{\RR}[2]{{#1}^{[#2]}}

\title{New Formulation for Coloring Circle Graphs\\
 and its Application\\
 to Capacitated Stowage Stack Minimization}
\author{Masato Tanaka \and Tomomi Matsui}

\begin{document}
\setcounter{tocdepth}{2}
\maketitle
\begin{center}
	\Large{Abstract}\\
\end{center}
\ \ 

A circle graph is a graph 
	in which the adjacency of vertices can be represented 
	as the intersection of chords of a circle. 
The problem of calculating the chromatic number 
	is known to be NP-complete, even on circle graphs.  
In this paper, we propose a new integer linear programming formulation
	for a coloring problem on circle graphs. 
We also show that the linear relaxation problem of our formulation
	finds the fractional chromatic number of a given circle graph.
As a byproduct, 
	our formulation gives a polynomial-sized
	linear programming formulation
	for calculating the fractional chromatic number 
	of a circle graph.

We also extend our result to a formulation for 
	a capacitated stowage stack minimization problem.
	
\section{Introduction}
 
This paper addresses problems of coloring circle graphs.
A circle graph is a graph
 in which the adjacency of vertices can be represented
  as the intersection of chords of a circle. 
It is well known
	that circle graphs and overlap graphs
	are of the same class (e.g., see~\cite{
		gavril1973algorithms,
		golumbic2004algorithmic}).

In~\cite{even1971queues},  
 	Even and Itai studied the problem of realizing 
 	a given permutation through networks of queues 
 	in parallel and through a network of stacks in parallel. 
The problem was translated into a coloring problem 
	of a circle graph
	 (see also~\cite{golumbic2004algorithmic}).
There are practical applications involving stack sorting
	including  
	assigning incoming trains~\cite{
		cornelsen2007track,
		demange2012online} 
	 or trams~\cite{blasum1999scheduling} to tracks 
	of a switching yard or depot; 
	parking buses in parking lots~\cite{gallo2001dispatching}; 
	and stowage planning 
	for container ships ~\cite{
		avriel2000container,
		wang2014stowage}.
K{\"o}nig  and L{\"u}bbecke~\cite{konig2008sorting}
	considered an algorithmic view towards stack sorting.
Stack and queue layouts of graphs related to colorations of circle graphs
	are discussed in~\cite{dujmovic2004linear}.

Even in the case of circle graphs, 
	the problems of finding the chromatic number~\cite{garey1980complexity}
	and clique covering number~\cite{keil2006approximating}
	are NP-complete.
Approximation algorithms for coloring circle graphs 
	are proposed in~\cite{
	vcerny2007coloring,
	supowit1985decomposing}.
In a survey~\cite{duran2014structural}, 
	Dur{\'a}n,  Grippo, and Safe			
	summarized structural results related to circle graphs 
	and presented some open problems.
Both the maximum clique problem 
	and maximum independent set problem
	have polynomial time algorithms when restricted 
	to circle graphs~\cite{
	apostolico1992new,
	vcenek2003maximum,
	gavril1973algorithms,
	rotem1981finding,
	valiente2003new}.

In this paper, we propose
	an integer linear programming formulation
	for coloring problems on circle graphs.
We also show that the linear relaxation problem of our formulation 
	finds the fractional chromatic number of a given circle graph.
For a general graph, 
	the problem of finding the fractional chromatic number
	is NP-complete~\cite{grotschel1981ellipsoid}.
Our proposal gives a polynomial-sized formulation 
	for fractional coloring problems on circle graphs.
	
The reminder of this paper is organized as follows. 
The next section presents some notations and definitions.
In Section~\ref{section:ILP}, 
	we propose a new formulation for coloring circle graphs.
We discuss a relation between the linear relaxation 
	of our formulation and fractional chromatic number 
	in Section~\ref{section:RelaxILP}. 
Section~\ref{section:CompExp} reports 
	our computational experiments.
In Section~\ref{section:StowageStack}, 
	we briefly discuss an extension of our formulation
	to a capacitated stowage stack minimization problem 
	with zero rehandle constraint.
Finally, Section~\ref{conclusion} makes some closing remarks.

\section{Notations and Definitions}

Let $G= (V, E)$ be an undirected graph 
	with a set of vertices $V$ and set of arcs $E$. 
A coloring of a graph is an assignment of a color to each vertex 
	such that all adjacent vertices are of a different color. 
The smallest number of colors needed to color a graph $G$ 
 is called its {\em chromatic number},  denoted by $\chi(G)$. 
The coloring problem has long been studied 
	and is known to be NP-complete 
	for general graphs~\cite{karp1972reducibility}.
An independent set is a subset of vertices in a graph 	
	such that no two are adjacent. 
The fractional chromatic number $\chi_f(G)$ 
 is the smallest positive number $k \in \mathbb{R}_+$ 
 for which there exists a probability distribution
  over the independent sets of $G$
  satisfying the following;  
 given an independent set $S$ drawn from the distribution,
 $\mbox{Pr} [v \in S] \geq 1/k \;\; (\forall v \in V)$.
Although its computation is NP-complete~\cite{grotschel1981ellipsoid}, 
	the fractional chromatic number of a general graph 
 can be obtained using linear programming
   (see Subsection~\ref{subsection:FC}). 
 
A clique is a subset of vertices such that
 its induced subgraph is complete. 
The clique number $\omega(G)$ of a given graph $G$
 is the number of vertices in a maximum clique in $G$.
It is known that
 	$\omega(G)\leq \vartheta(\overline{G})\leq \chi_f(G)\leq \chi(G)$,
 	where 
 	$\vartheta(\overline{G})$ denotes 
 	the Lova\'{s}z number~\cite{lovasz1979shannon,knuth1994sandwich}
 	of a given graph $G$. 
A pentagon graph (5-cycle) $C_5$, 
	which is an example of a circle graph,  satisfies 
 $(\omega(C_5), \vartheta(\overline{C_5}), \chi_f(C_5),  \chi(C_5))=
  (2, \sqrt{5}, 2.5, 3)$.

\begin{figure}[h]
\centering
\includegraphics[width=12cm]{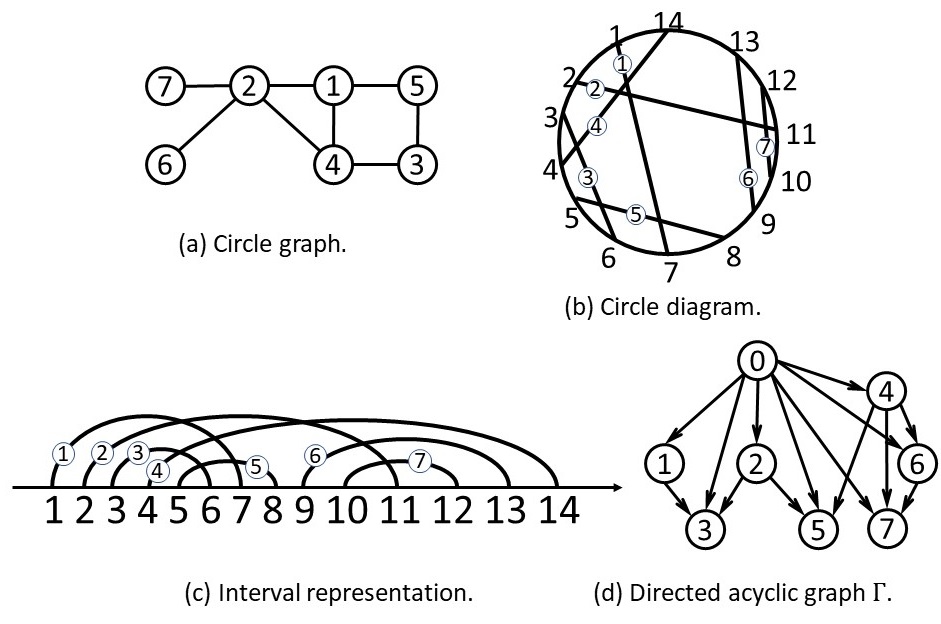}
\caption{An example of circle graph and related diagram.}\label{ex1}
\end{figure}

A {\em circle graph} 
 is a graph in which the adjacency of vertices
  can be represented as the intersection of chords of a circle. 
The circle and chords
 corresponding to a given circle graph $G$
  are called a {\em circle diagram} of $G$. 
Hereinafter, 
  we assume that 
  terminal points of chords in a circle diagram
  are mutually distinct. 
Figure~\ref{ex1}~(a) and~(b) show an example 
 of a circle graph and its corresponding circle diagram, respectively.
%

It is well known
	that circle graphs and overlap graphs
	are of the same class (e.g., see~\cite{
		gavril1973algorithms,
		golumbic2004algorithmic}).
A graph is an {\em overlap graph} 
	if its vertices are intervals on a line
	such that two vertices are adjacent 
	if and only if the corresponding intervals partially overlap
	(that is, they have non-empty intersection), 
	 but neither contains the other.
It is easy to construct a set of intervals  
	representing a given circle graph (overlap graph) 
	from a corresponding circle diagram
	by a simple transformation:
	cutting the circumference
	of the circle at some point $p$ 
	that is not an endpoint of a chord 
	and unfolding it at that point
     (e.g., see Section~11.3 of~\cite{golumbic2004algorithmic}).
Hereinafter, 
	we assume that an input of a given circle graph $G=(V, E)$
	is a corresponding  {\em interval representation}   
	${\cal I}(G) =\{ I(j)  \subseteq \mathbb{R} \mid j \in V\}$, 
	where $I: j \mapsto [l_j, r_j]$.
We also assume that all the 
	terminal points of intervals in ${\cal I}(G)$ are mutually distinct. 
Here, we note that an interval representation 
	of a given circle graph is not unique.
Figure~\ref{ex1}~(c) shows an interval representation 
 of the circle graph in Figure~\ref{ex1}~(a).
 	
\begin{figure}[h]
\centering
\includegraphics[width=4cm]{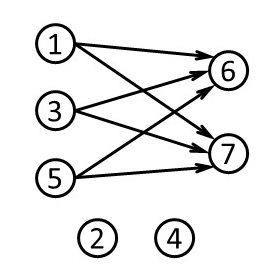}
\caption{Partially ordered set corresponding to
	the interval representation in Figure~\ref{ex1}~(c).
	There is an arrow from $i$ to $j$ if and only if $i \preceq j$.}\label{exPO}
\end{figure}

Given an interval representation ${\cal I}(G)$ 
 of a circle graph $G=(V, E)$, 
 we introduce a partial order $\preceq$ defined on the vertex set $V$. 
For any pair of vertices $i, j \in V$, 
	we define $i \preceq j$ if and only if  
	either $i=j$ or $r_i \leq l_j$ holds,
	where $I(i)=[l_i, r_i]$ and $I(j)=[l_j, r_j]$.
Obviously, $(V, \preceq) $ is a partially ordered set.
Although every chain of $(V, \preceq) $ is an independent set 
	of $G$, the converse implication does not hold.
Figure~\ref{exPO} shows a partially ordered set 
 corresponding to the interval representation in Figure~\ref{ex1}~(c).

\section{Integer Linear Programming Formulation} \label{section:ILP}
\newcommand{\CH}{\mbox{\rm Ch}}
\newcommand{\PRT}{\mbox{\rm Prt}}

Given a circle graph $G=(V, E)$ 
	and  corresponding interval representation 
	${\cal I}(G)$, 
	we introduce a directed graph $\Gamma$ as follows.
The vertex-set of $\Gamma$ is defined by $V \cup \{0\}$,
	where $0$ is an artificial vertex called a {\em root}.
The arc-set of $\Gamma$, denoted by $A$, is defined by
\[
	A=\{(0, i) \mid i \in V\} \cup \{(i, j) \mid I(i) \supsetneq I (j)\}.
\]
The above definition implies that
	$\Gamma$ is acyclic.
Figure \ref{ex1}~(d) shows a directed acyclic graph $\Gamma$
 defined by the interval representation in~(c).
	
An arc subset $T \subseteq A$ is called an {\em arborescence}
	if and only if
	 $|T|=|V|$ and
	 each vertex $i \in V$ has a unique incoming-arc in $T$.
When a given arborescence $T$ has an arc $(i, j)$, 
	we say that $j$ is a {\em child} of $i$ and 
	$i$ is a (unique) {\em parent} of $j$  with respect to $T$.
For any arborescence  $T$ and a vertex  $i \in V\cup \{0\}$,
	$\CH (T, i)$ denotes the set of children 
	of $i$ with respect to $T$.

\begin{figure}[h]
\centering
\includegraphics[width=12cm]{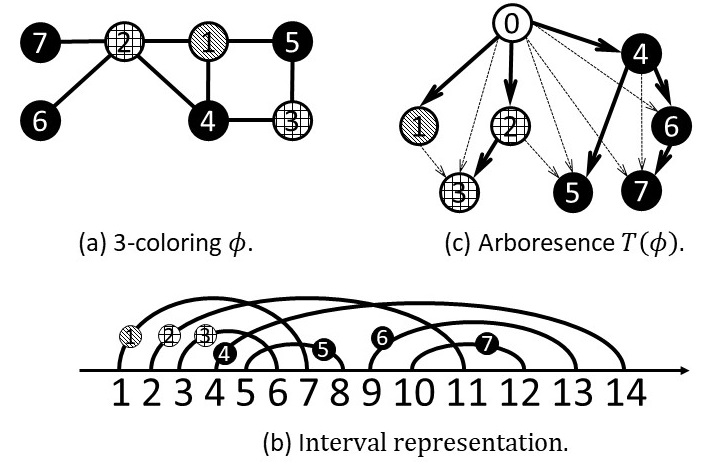}
\caption{An example of a 3-coloring 
	and its corresponding arborescence.}\label{ex2}
\end{figure}

In the following, 
	we associate each coloring with an arborescence on $\Gamma$.
Let $\phi: V \rightarrow \{1, 2, \ldots , c\}$ be a $c$-coloring of  $G$.
For each vertex $j \in V$, 
  	we define a parent of $j$ with respect to $\phi$, 
	denoted by $\PRT (\phi, j)$, 
	as follows:  
	if $V'= \{i \in V \mid \phi (i)=\phi (j), I(i) \supsetneq I(j) \}$
	is empty, 
	then we define $\PRT (\phi, j)=0$ (root); 
	else, $\PRT (\phi, j) $ denotes a vertex in $V$
	corresponding to a unique (inclusion-wise) minimum interval 
	 in $V'$.
Given a coloring $\phi$ of $G$,
	$T(\phi)$  denotes an arborescence 
	$\{(\PRT (\phi, j), j) \in A \mid j \in V\}$. 
Figure~\ref{ex2}~(a) shows a 3-coloring
	and  corresponding arborescence in $\Gamma$.
	
\begin{lemma} \label{arboriff}
Let $T$ be an arborescence of  $\Gamma$.
Then, there exists a  $c$-coloring $\phi$ of a given circle graph $G$ 
   satisfying $T=T(\phi)$
	if and only if 
\begin{description}

\item[C1:] for each $i \in V$, 	$\CH (T, i)$ 
 is a chain of $(V, \preceq)$ or the empty set  and
\item[C2:] 
 the size of every antichain of $(V, \preceq)$ 
  contained in $\CH (T, 0)$ is less than or equal to $c$.
\end{description}
\end{lemma}

\noindent
Proof.   
%
%
It is obvious that 
	the size of a minimum chain cover (partition) of a poset is 
 	greater than or equal to the size of a maximum antichain. 
Thus,  the definition of $T(\phi)$ implies that
	if there exists a $c$-coloring $\phi$ of $G$ 
	satisfying $T=T(\phi)$, 
	then $T$ satisfies C1 and C2. 

We show the converse implication. 
In the following,  we construct a \mbox{$c$-coloring}  
	from an arborescence $T$ satisfying C1 and C2. 
Because $T$ satisfies C2, 
	Dilworth's theorem~\cite{dilworth2009decomposition} implies that
  there exists a set of (at most) $c$ chains of $(V, \preceq)$ 
   partitioning $\CH (T,0)$.  
Because every chain of $(V, \preceq)$ is 
	an independent set of a given circle graph, 
	we obtain a \mbox{$c$-coloring}  of 
	a sub-graph of $G$ induced by $\CH (T,0)$
	by assigning a color to each chain. 
For each vertex $i \in \CH (T,0)$, 
	we assign a color of $i$ to all the descendants of $i$
	with respect to $T$.  
Denote the map (coloring) obtained above by $\phi$.
We only need to show that
	if $\phi (j)=\phi (j')$ and $j \neq j'$, 
	then vertices $j$ and $j'$ are non-adjacent 
	on a given circle graph $G$.
When $I(j) \subseteq I(j')$ or $I(j') \subseteq I(j)$, 
	the non-adjacency is obvious. 
Otherwise, let $r'$ be a unique lowest common ancestor
	of $j$ and $j'$ with respect to $T$.
We denote a child of $r'$ 
	that is an ancestor of $j$ (or $j'$) by $i$ (or $i'$), respectively.  
If $r'\neq 0$, then C1 directly implies that 
	$I (i) \cap I(i') =\emptyset$.
When $r'=0$, $\phi (i)=\phi (j)=\phi(j')=\phi (i')$ implies that 
	$i$ and $i'$ are contained in a mutual chain in $\CH (T,0)$ 
	and thus  $I (i) \cap I(i') =\emptyset$.
From the above, 
	$I(j) \subseteq I(i),$ $I(j') \subseteq I(i'),$ and 
	$I (i) \cap I(i') =\emptyset$ hold. 
As a consequence,
  we obtain  the non-adjacency of $j$ and $j'$,
  because $I(j) \cap I(j') =\emptyset$.
\QED

\bigskip 

Let $P \subseteq \mathbb{R}$ be a set of (positions of) terminal points 
	of intervals in ${\cal I}(G)$.
Recall that 
	terminal points of intervals in ${\cal I}(G)$ are mutually distinct
	and thus $|P|=2|V|$. 
Let $M=(m_{pi})$ be a 0-1 matrix
	 whose entries are indexed by $P \times V$, satisfying 
\[
	m_{pi}=\left\{
	\begin{array}{ll}
		1 & (\mbox{if } p \in I(i) ), \\
		0 & (\mbox{otherwise}). 
	\end{array}
	\right.
\]
(Here, we note that $M$ is a {\em clique matrix} 
	(Section~3.4 of~\cite{golumbic2004algorithmic})
	 of an interval graph corresponding 
	 to the set of intervals ${\cal I}(G)$.) 
Obviously, 
	$M$ is an antichains-versus-vertices incidence matrix 
	of poset $(V, \preceq)$.
In addition, it is easy to see that 
	all the maximal antichains are included.
The definition of $M$ directly implies 
	the following lemma, which characterizes
	the size of a maximum antichain.
	
\begin{lemma} \label{antichain-cM}
For any vertex subset $\widetilde{V} \subseteq V$, 
	the size of a maximum antichain of $(V, \preceq)$ 
  contained in $\widetilde{V} $ is equal to the maximum components 
  of vector $M \widetilde{\bm x}$,
where $\widetilde{\bm x}$ is 
	the (fixed) characteristic vector of $\widetilde{V}$. 
\end{lemma} 

\noindent
Proof is omitted.
%

\bigskip 

Now, we give our formulation for a circle graph coloring problem.
For any vertex $i \in V\cup \{0\}$, 
	we define a vertex subset
	$\RR{V}{i}=\{j \in V \mid (i, j) \in A\}$.
We define $V^{\bullet}=\{i \in V \mid \RR{V}{i}\neq \emptyset\}$.
Here, we note that 
	$\RR{V}{0}=V$ and  $0 \not \in V^{\bullet}$  hold. 
For each arc $(i, j) \in A$, we introduce a 0-1 variable $x^i_j$.
The vector of all 0-1 variables is denoted by ${\bm x} \in \{0, 1\}^A$. 
For any vertex $i \in V^{\bullet} \cup \{0\}$,  
  $\RR{\bm x}{i}$ denotes a subvector of ${\bm x}$ indexed by 
  arcs emanating from $i$,  and
	$\RR{M}{i}$ denotes a submatrix of $M$ consisting of 
	column vectors of $M$ indexed by  $\RR{V}{i}$. 
Then, we have the following.

\begin{lemma}\label{01feasible}
Given a vector ${\bm x} \in \{0, 1\}^A$ 
 and positive integer $c$, the constraints
\[
\begin{array}{ll}
	M\RR{\bm x}{0} \leq \ c{\bm 1},\\  
 	\RR{M}{i}\RR{\bm x}{i} \leq {\bm 1}\    & (\forall i\in V^{\bullet}), \\
\displaystyle  \sum_{i: (i, j)\in A} x^i_j = 1\  & (\forall j\in V) 
\end{array}
\]
are satisfied if and only if
	$T=\{(i, j) \in A \mid x^i_j =1 \}$ is an arborescence 
	satisfying conditions C1 and C2.
\end{lemma}

\noindent
Proof.  
Let $T$ be an arborescence  satisfying conditions C1 and C2.
We set a vector  ${\bm x} \in \{0, 1\}^A$ as the characteristic 
	vector of $T$.
Then, it is obvious that ${\bm x}$ satisfies the above constraints.

Now, assume that ${\bm x} \in \{0, 1\}^A$ 
	and a positive integer $c$ satisfy the above constraints. 
We define $T=\{(i, j) \in A \mid x^i_j =1 \}$.
Then, constraints $\sum_{i: (i, j)\in A} x^i_j = 1\   (\forall j\in V) $
	directly imply  that $T$ is an arborescence of $\Gamma$.
From Lemma~\ref{antichain-cM}, 	
	 the inequality  $M \RR{\bm x}{0} \leq \ c{\bm 1} $ implies that
	 $T$ satisfies condition~C2. 
Similarly, Lemma~\ref{antichain-cM} and  	
	$\RR{M}{i}\RR{\bm x}{i} \leq {\bm 1}\  (\forall i\in V^{\bullet})$
	imply that for any $i \in  V^{\bullet},$ 
	the size of a maximum antichain in $\CH (T, i)$ is 
	less than or equal to $1$.
From Dilworth's theorem,   
	$\CH (T, i)$ becomes a chain (or the empty set).
For any $i \in V\setminus V^{\bullet}$, 
	$\CH (T, i)=\emptyset$.
Thus, $T$ satisfies condition C1.
\QED

\medskip

The above lemma directly implies the following 
	formulation for a circle graph coloring problem:

\begin{subequations}
	\label{coloring}
	\begin{alignat}{2}
		{\rm CG:} \: {\rm min.}\ & c \nonumber \\
		{\rm s.t.}\ & M\RR{\bm x}{0} \leq \ c{\bm 1}, 
			&&  \nonumber \\ 
		& \RR{M}{i}\RR{\bm x}{i} \leq {\bm 1}\   
			&& (\forall i\in V^{\bullet}),   \nonumber \\
		& \sum_{i: (i, j)\in A} x^i_j = 1\  
			&& (\forall j\in V),  \nonumber \\
		& x^i_j\in \{0, 1\}\ 
			&& (\forall (i,  j) \in A), \nonumber \\
		& c\in \mathbb{Z}_+.  \nonumber 
	\end{alignat}
\end{subequations}
Lemma~\ref{01feasible} directly implies the following.

\begin{theorem}
A pair  
 $(\widehat{\bm x}, \widehat{c}) \in \{0, 1\}^A \times  \mathbb{Z}_+$
	is optimal to CG
	if and only if
	$\widehat{c}=\chi (G)$ and there exists 
	a $\widehat{c}$-coloring $\phi$ 
	satisfying $T(\phi) =\{ (i, j) \in A \mid \widehat{x}^i_j=1\} $.
\end{theorem}

\noindent
Proof.  
Lemma~\ref{arboriff} implies that
	a pair  $({\bm x}, c)  \in \{0, 1\}^A \times  \mathbb{Z}_+$
	is feasible to CG, if and only if, 
	$T=\{ (i, j) \in A \mid x^i_j=1\} $ and $c$ satisfies conditions 
	C1 and C2.
Thus, the optimal value of CG is equal to $\chi (G)$.
We can construct a $\chi (G)$-coloring $\phi$ 
	from an optimal solution of CG
	by applying a technique described in the proof of Lemma~\ref{arboriff}.
The inverse implication is clear.
\QED


\section{Linear Relaxation of ILP formulation}\label{section:RelaxILP}

In this section, 
we show that the linear relaxation problem of our formulation (CG)
	finds the fractional chromatic number of a given circle graph.

\subsection{Maximum Weight Independent Set Problem}

In this subsection, 
	we discuss a maximum weight independent set problem
	defined on a given circle graph $G=(V,E)$ 
	with a given vertex weight function $w:V \rightarrow \mathbb{R}$
	(incidentally, negative vertex weights are permitted).  
For an artificial vertex $0$, we define $w(0)=0$.
We propose a linear programming formulation 
	of the problem based on a dynamic programming technique~\cite{
	vcenek2003maximum,
	felsner1997trapezoid,
	gavril1973algorithms}.
Our linear programming formulation plays an important role 
	in the next subsection.

A maximum weight independent set problem finds 
	an independent set $S$ of $G$
	that maximizes the weight $\sum_{i \in S}w(i)$.
Throughout this section, 
	we assign a linear ordering on the vertex set
	by setting $V=\{1,2,\ldots, n\}$
	such that if $I(i) \supseteq I(j)$, then $i \leq j$.

\begin{figure}[h]
\centering
\includegraphics[width=12cm]{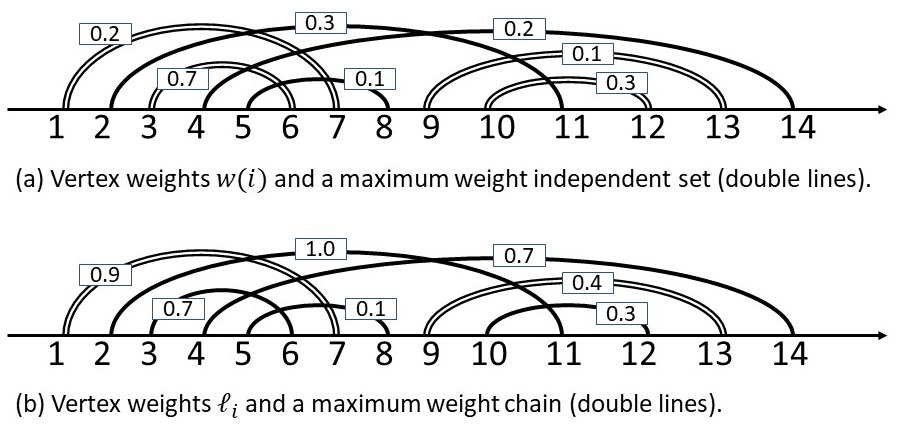}
\caption{An example solution of recursive formula~(\ref{RF}).}\label{ex3}
\end{figure}

For any vertex $i \in V^{\bullet} \cup \{0\}$,  
	$\RR{G}{i}$ denotes the subgraph of $G$ 
	induced by vertex subset $\RR{V}{i}$
	(note that $i \not \in \RR{V}{i}$).
Every maximum weight independent set $S \subseteq V$ satisfies 
	the following:
	$\forall i \in S \cap V^{\bullet}$, 
	$S \cap \RR{V}{i}$ is a maximum weight independent set 
	in $\RR{G}{i}$.
Here, we introduce vertex weights defined by 
\[
	\ell_i=\left\{
		\begin{array}{ll} \displaystyle 
			w(i)+\max \left\{ \left. \sum_{i \in V'} w(j) \right| \
			\begin{array}{l}
				V' \mbox{ is an independent} \\
				\mbox{ set of } \RR{G}{i} 
			\end{array} 
			\right\}
				& (\forall i \in V^{\bullet}), \\ 
			w(i) & (\forall i  \in V \setminus V^{\bullet}).
		\end{array} \right.
\]
For any vertex subset $S \subseteq V$, 
	$\max S$ denotes a set of vertices corresponding to
	(inclusion-wise) maximal intervals in $\{ I(i) \mid i \in S \}$.
It is clear that
	if $S$ is an independent set of $G$, 
	then  $\max S$ is a chain of $(V, \preceq)$.
This property implies that
	the weight of a maximum weight independent set
	with respect to  $(w(i) \mid i \in V)$ 
	is equal to the weight of a maximum weight chain
	 (of   $(V, \preceq)$)
	with respect to $(\ell_i \mid i \in V)$. 
By applying the above idea recursively,  
	it is easy to see that $( \ell_i \mid i \in V \cup \{0\}) $ satisfies
	the following formula 
\begin{equation} \label{RF}
		\ell_i=\left\{
		\begin{array}{ll}
			w(i) +  \max \left\{  \left. \displaystyle 
				\sum_{j \in V'} \ell_j  \right|
					\begin{array}{l}
						V' \subseteq  \RR{V}{i}, \\
						V' \mbox{ is a chain of } (V, \preceq)
					\end{array}	
				\right\} &
				(\forall i \in V^{\bullet} \cup \{0\}), \\
			w(i) & (\forall i \in V\setminus  V^{\bullet}), 
		\end{array}\right.
\end{equation}
	where we define $w(0)=0$, and $\ell_0$ denotes 
	the weight of a maximum weight independent set
	of $G$ with respect to  $(w(i) \mid i \in V)$. 
Because the vertex set $V=\{1,2,\ldots, n\}$
	satisfies  ``if $I(i) \supseteq I(j)$, then $i \leq j$,'' 
	the above formula calculates 
	$(\ell_n, \ell_{n-1}, \ldots , \ell_1, \ell_0)$
	sequentially.
Figure~\ref{ex3} shows an example solution 
	of recursive formula~(\ref{RF}).

The following lemma summarizes the above discussion.
 
\begin{lemma} \label{lemma:recursive}
The solution of recursive formula~(\ref{RF}) satisfies that 
	$\ell_0$ is equal to the weight of a maximum weight independent set
	with respect to $(w(i) \mid i \in V)$.
\end{lemma}

Next, we describe a linear programming formulation 
	of a subproblem appearing in~(\ref{RF}).
\begin{lemma} \label{lemma:consecutive1}
Let ${\bm \ell}=(\ell_j \mid j \in V)$ represent given vertex weights. 
For any $i \in V^{\bullet} \cup \{0\}$, 
\begin{equation} \label{maxchain}
 \max \left\{  \left.  
		\sum_{j \in V'} \ell_j  \right|
					\begin{array}{l}
						V' \subseteq  \RR{V}{i}, \\
						V' \mbox{ is a chain of } (V, \preceq)
					\end{array}	
		\right\}
= \max \left\{ \left.
		  \sum_{j \in \RR{V}{i}} \ell_j x^i_j \right|
		  		\begin{array}{l}
		  			\RR{M}{i}\RR{\bm x}{i} \leq {\bm 1}, \\
		  			\RR{\bm x}{i} \geq {\bm 0}
				\end{array}
		  \right\},
\end{equation}
where $\RR{\bm x}{i} $ is a vector of continuous variables
	indexed by $\RR{V}{i}$.
\end{lemma}

\noindent
Proof. 
In the following, ${\rm LC}_i ({\bm \ell})$ denotes
	the linear programming problem on the
	right-hand-side of~(\ref{maxchain}).
First, we show that ${\rm LC}_i ({\bm \ell})$
	 has an optimal 0-1 vector solution.
We say that a matrix has the consecutive 1's property (for columns)
	if and only if 1's occur consecutively in each column. 
Clearly, the matrix $\RR{M}{i}$ has the consecutive 1's property~\cite{
	fulkerson1965incidence}
(for columns) and is totally unimodular.
The total unimodularity implies that
	a feasible region 
	$\{ \RR{\bm x}{i} \mid \RR{M}{i} \RR{\bm x}{i} \leq {\bm 1}, 
		\RR{\bm x}{i} \geq {\bm 0} \}$
	is a 0-1 polytope~\cite{
		hoffman1956integral,
		hoffman1974generalization}.
Therefore, ${\rm LC}_i  ({\bm \ell})$
	 has an optimal 0-1 vector solution, 
	 denoted by $\RR{\widehat{\bm x}} {i}$.
Lemma~\ref{antichain-cM} and  	
	$\RR{M}{i}\RR{\widehat{\bm x}}{i} \leq {\bm 1} $ 
	imply that  
	the size of a maximum antichain contained in 
	$\widehat{V}=\{j \in \RR{V}{i} \mid \widehat{x}^i_j=1 \}$ is 
	less than or equal to $1$.
From Dilworth's theorem,   
	$\widehat{V}$ becomes a chain (or the empty set).	  
\QED

\medskip

For any $i \in V^{\bullet} \cup \{0\}$,
	the dual of LC$_i ({\bm \ell})$, 
	denoted by DLC$_i ({\bm \ell})$, is 
	\begin{alignat*}{2}
	{\rm DLC}_i ({\bm \ell}){\rm :}\;
	{\rm min.}\ & \sum_{p \in P} y^i_p  \\
	{\rm s.t.}\ 
		& \sum_{p \in I(j)} y^i_p \geq \ell_j\    && (\forall j\in \RR{V}{i}),  \\
		& y^i_p \geq 0 && (\forall p \in P).
	\end{alignat*}

We substitute DLC$_i ({\bm \ell})$ 
	for the maximum weight chain problem
	in~(\ref{RF})
	to obtain the following recursive formula:
\begin{equation} \label{LPRF}
\ell_i=\left\{
	\begin{array}{ll}
		\displaystyle
		w(i)+\min \left\{  \sum_{p \in P} y^i_p  \left|  
		\begin{array}{ll}
			\displaystyle
			\sum_{p \in I(j)} y^i_p \geq \ell_j & (\forall j\in \RR{V}{i}), \\
			y^i_p \geq 0  & (\forall p \in  P)
		\end{array} \right. \right\}
			& (\forall i \in  V^{\bullet} \cup \{0\}),\\
		w(i) &  (\forall i \in V\setminus V^{\bullet}).
	\end{array} \right.
\end{equation}

\noindent
Lemmas~\ref{lemma:recursive} and \ref{lemma:consecutive1} 
	and the strong duality theorem
 	directly imply the following.

\begin{theorem}
The solution of recursive formula~(\ref{LPRF}) satisfies that 
	$\ell_0$ is equal to the weight of a maximum weight independent set
	with respect to vertex weights ${\bm w}=(w(i) \mid i \in V)$.
\end{theorem}

\noindent
Considering all constraints appearing in~(\ref{LPRF}), 
	 we construct the following linear programming problem:
\begin{subequations}
	\label{ISD}
	\begin{alignat}{2}
	{\rm ISD}({\bm w}):\;
	{\rm min.}\ & \ell_0=\sum_{p \in P}y^0_p  \\
	{\rm s.t.}\ 
		& \ell_i=w(i)+\sum_{p \in P} y^i_p \;\;\;
		 &&(\forall i \in V^{\bullet}),  \label{eliminatew1} \\
		& \ell_i=w(i) 
		 &&(\forall i \in V \setminus V^{\bullet}), \label{eliminatew2} \\
		& \sum_{p \in I(j)} y^i_p \geq \ell_j\    
			&& (\forall (i,j) \in A),  \\
		& y^i_p \geq 0 && (\forall (i,p) \in (V^{\bullet} \cup \{0\}) \times P),
	\end{alignat}
\end{subequations}
where $(\ell_0, \ell_1, \ldots , \ell_n)$ and 
	 $(y^i_p  \mid  (i,p) \in (V^{\bullet} \cup \{0\}) \times P)$ 
	 are vectors of continuous variables. 

\newcommand{\optell}[1]{\ell^*_{#1}}
\newcommand{\optelll}{\ell^*}
\newcommand{\opty}{{y^*}}

\begin{theorem}
The optimal value of ISD$(\bm w)$ is equal 
	to the weight of a maximum weight independent set
	with respect to vertex weights ${\bm w}=(w(i) \mid i \in V)$.	
\end{theorem}

\noindent
Proof. 
Let	$(\widetilde{\bm \ell}, \widetilde{\bm y})$ 
	be a solution of~(\ref{LPRF})
	and $({\bm \optelll}, {\bm \opty})$  
	be an optimal solution of ISD$(\bm w)$.
Because $(\widetilde{\bm \ell}, \widetilde{\bm y})$ 
	is feasible to ISD$(\bm w)$,
	$\widetilde{\ell}_0 \geq \optell{0}$ holds.


For the reminder of this proof, we show the inequality 	
	$\widetilde{\ell}_0 \leq \optell{0}$.
Here, we note that 
	we assign a linear ordering on the vertex set
	by setting $V=\{1,2,\ldots, n\}$
	such that if $I(i) \supseteq I(j)$, then $i \leq j$.
We show that $\widetilde{\ell}_j \leq \optell{j}$ 
	for each $j\in \{n, n-1, \ldots , 0\} $ by induction on $j$.
Clearly, vertex $n  \in V\setminus V^{\bullet}$, and thus 
	$\optell{n}= w(n)=\widetilde{\ell}_n$.
Assume that $\widetilde{\ell}_{j}\leq \optell{j}$ 
	for each $j \in \{n, n-1, \ldots , i+1\}$.
When $i  \in V\setminus V^{\bullet}$, we obviously have 
 	$\optell{i}= w(i)=\widetilde{\ell}_i$.
We consider the case that $i \in V^{\bullet} \cup \{0\}$. 	
It is obvious that for any $i \in V^{\bullet} \cup \{0\}$, 
	the subvector $(\widetilde{y}^i_p)_{p \in P}$ of $\widetilde{\bm y}$ 
	is optimal to problem  DLC$_i (\widetilde{\bm \ell})$.
The subvector $({\opty}^i_p)_{p \in P}$ of ${\bm \opty}$ 
	is feasible to problem DLC$_i ({\bm \ell}^*)$.
The induction hypothesis implies that the feasible region of 
	 DLC$_i (\widetilde{\bm \ell})$ includes that
	 of DLC$_i ({\bm \ell}^*)$.
The subvector $({\opty}^i_p)_{p \in P}$ of ${\bm \opty}$
	is feasible to   DLC$_i (\widetilde{\bm \ell})$, 
	and the corresponding objective value satisfies 
\[
	\sum_{p \in P} \opty^i_p  
	\geq  (\mbox{optimal value of DLC}_i  (\widetilde{\bm \ell})  ) 
	= \sum_{p \in P} \widetilde{y}^i_p. 
\]
Thus, we obtain 
\[
	\optell{i} =w(i)+ \sum_{p \in P} \opty^i_p 
	\geq w(i)+  \sum_{p \in P} \widetilde{y}^i_p 
	=\widetilde{\ell}_i,
\]
where we let $w(0)=0$ for simplicity.  

From the above discussion, 
	we have shown that $\optell{0}=\widetilde{\ell}_0$. 
\QED

\bigskip

\subsection{Fractional Coloring Problem}\label{subsection:FC}

In this subsection,
  we discuss the fractional coloring problem.
Given an  undirected graph  $G=(V,E)$,  
	$F$ denotes the incidence matrix 
	of independent sets of $G$.
The rows of $F$ are indexed by $V$,
  the columns of $F$ are indexed by 
    all the independent sets of $G$,
  and each column vector is 
   the incidence vector (characteristic vector)
   of a corresponding independent set.
The fractional coloring problem is defined by 
\[  \min \{ {\bm 1}^{\top} {\bm q}   
      \mid F {\bm q} = {\bm 1},  {\bm q} \geq {\bm 0} \},  
\]
	where the variable vector ${\bm q}$ is indexed by 
	all the independent sets in $G$, 
	and ${\bm 1}$ denotes the all-ones vector.
The optimal value of the above problem is called 
	the {\em fractional chromatic number}
	and is denoted by $\chi_f (G)$.
Generally, 
  the above linear programming problem 
  has an exponential number of variables.
The dual of the above problem is
\[
	\max \{ {\bm w}^{\top} {\bm 1} \mid
	{\bm w}^{\top} F \leq {\bm 1}^{\top}  \},
\]  
which finds a  vertex weight $w: V \rightarrow \mathbb{R}$ 
	maximizing the total sum $ {\bm w}^{\top} {\bm 1}$ 
	subject to the constraint that the weight of every independent set 
	(with respect to ${\bm w}$) is less than or equal to 1.

Let us discuss the case that a given graph $G$ is a circle graph.
Given a vertex weight function $w: V \rightarrow \mathbb{R}$,
 	the weight of every independent set 
	(with respect to ${\bm w}$) 
	is less than or equal to 1 if and only if 
 	the minimization problem ISD$(\bm w)$ has 
 	a feasible solution whose objective value
 	is less than or equal to~1.
Then, the linear programming problem 

\begin{eqnarray}
\mbox{max.} &&    {\bm w}^{\top} {\bm 1}  \nonumber \\
\mbox{s.t.} && \sum_{p \in P} y^0_p \leq 1, \label{FCP} \\
				 && \mbox{with the constraints of the ISD problem},  \nonumber
\end{eqnarray}
	gives a formulation for the fractional coloring problem
	on a given circle graph,
	where $(w(i) \mid i \in V)$, 
	 $(\ell_i \mid i \in V)$, and 
	 $(y^i_p  \mid  (i,p) \in (V^{\bullet} \cup \{0\}) \times P)$ 
	 are vectors of continuous variables.
Here, we note that  $(w(i) \mid i \in V)$ 
	is a given vector of vertex weights in ISD$(\bm w)$
	and is a variable vector in the above problem.

We eliminate variables  $(w(i) \mid i \in V)$ 
	by applying equalities~(\ref{eliminatew1}) and~(\ref{eliminatew2}).
Then, the objective function becomes
\begin{eqnarray*}
 {\bm w}^{\top} {\bm 1}&=&\sum_{i \in V^{\bullet}} w(i) + \sum_{i \in V\setminus V^{\bullet}} w(i) \\
 &=& \sum_{i \in V^{\bullet}} (\ell_i - \sum_{p \in P} y^i_p ) 
 	+ \sum_{i \in V\setminus V^{\bullet}} \ell_i 
 	= \sum_{j \in V}  \ell_j - \sum_{i \in V^{\bullet}} \sum_{p \in P} y^i_p.
\end{eqnarray*}
The obtained problem~(\ref{FCP}) transforms into 
	\begin{alignat*}{2}
	{\rm FCP:}\;
	{\rm max.}\ & \sum_{j \in V}  \ell_j - \sum_{i \in V^{\bullet}} \sum_{p \in P} y^i_p \\
	{\rm s.t.}\ 
		& \sum_{p \in P} y^0_p \leq 1, \\
		& \sum_{p \in I(j)} y^i_p \geq \ell_j   \    
			&& (\forall (i,j) \in A),  \\
		& y^i_p \geq 0 && (\forall (i,p) \in (V \cup \{0\}) \times P), 
	\end{alignat*}
where  $(\ell_i \mid i \in V)$  and 
	 $(y^i_p  \mid  (i,p) \in (V^{\bullet} \cup \{0\}) \times P)$ 
	 are vectors of continuous variables.

It is easy to check that 
	FCP is the dual of the linear relaxation problem of CG
	obtained by substituting $x^i_j \geq 0$ and $c \geq 0$ 
	for $x^i_j \in \{0, 1\}$ and $c \in  \mathbb{Z}_+$, 
	respectively. 
Summarizing the above discussion, 
	we obtain the following theorem.
	
\begin{theorem}
Let LR be a  linear relaxation problem of CG
	obtained by substituting $x^i_j \geq 0$ and $c \geq 0$ 
	for $x^i_j \in \{0, 1\}$ and $c \in  \mathbb{Z}_+$, 
	respectively. 
Then, the optimal value of LR 
	is equal to the fractional chromatic number 
	of a given circle graph.
\end{theorem}
 

\section{Computational Experiments} \label{section:CompExp}

In our experiments, we compared the computational time required 
  to find an optimal solution of our  formulation CG, 
  a classical coloring problem formulation
\begin{subequations}
	\begin{alignat}{2}
		{\rm CL: min.}\ &\sum^{C}_{c=1}\ y_c,  \nonumber \\
		{\rm s.t.}\ & x_{ic} \leq y_c\ 
			&&(\forall i\in V, \forall c \in \{1, 2, \ldots , C\}),  \nonumber \\
		& x_{ic} + x_{i'c} \leq 1\ 
			&&(\forall c\in \{1, 2, \ldots , C\}, \forall \{i,i'\}\in E), \nonumber \\
		& \sum^{C}_{c=1} x_{ic}\geq 1\ 
			&& (\forall i\in V),   \nonumber \\
		& x_{ic}\in \{0, 1\}\ 
			&&(\forall i\in V, \forall c \in \{1, 2, \ldots , C\}),  \nonumber \\
		& y_c\in \{0, 1\}\ 
			&& (\forall c \in \{1, 2, \ldots , C\}),  \nonumber 
	\end{alignat}
\end{subequations} 
  and an asymmetric representative  formulation~\cite{
  	campelo2008asymmetric}
\begin{subequations}
	\begin{alignat}{2}
		{\rm AS: min.}\ &\sum_{i\in V}\ x_{ii}  \nonumber \\ 
		{\rm s.t.}\ & x_{ij}=x_{ji}=0\ 
			&& (\forall \{i,j\}\in E),  \nonumber \\
		& x_{ji} = 0 \ 
			&& (\forall i, \forall j \in V, \  i < j ),  \nonumber \\
		& x_{ij}+x_{ik}\leq x_{ii}\ 
			&&(\forall \{i, j, k\} \subseteq V,  \{j, k\}\in E),  \nonumber \\
		& \sum_{i\in V} x_{ij} = 1\ 
			&&(\forall j\in V),  \nonumber \\
		& x_{ij}\leq x_{ii}\ 
			&&(\forall i, \forall j\in V),   \nonumber \\
		& x_{ij} \in \{0,1\} 
			&& (\forall (i,j)\in V^2),   \nonumber 
	\end{alignat}
\end{subequations}
	where the vertex set $V=\{1, 2, \ldots , n\}$
	satisfies $\mbox{deg}(1) \geq \mbox{deg}(2) \geq 
	\cdots \geq \mbox{deg}(n)$ 
	($\mbox{deg}(v)$ denotes the degree of vertex $v \in V$). 
We set the constant $C$ in the classical formulation
	to the number of colors required in a coloring obtained 
	by the First Fit heuristic~(e.g., see \cite{aslan2016performance}).
All the experiments were conducted on a PC running 
	the Windows 10 Pro operating system 
	with an Intel(R) Core(TM) i7-7700 @3.60GHz processor and 
	32 GB RAM.
All instances were solved using 
	CPLEX 12.8.0.0 implemented in Python 3.6.5 and
	NumPy 1.17.2. 
 
We generated instances of circle graphs as follows. 
First, we randomly shuffled the numbers $\{1, 2, \ldots , 2|V|\}$
	using the ``random.shuffle()'' command 
	of the NumPy Python module. 
We repeatedly removed the first two numbers $x, y$
	from the shuffled sequence
	and added a non-empty interval $[x, y]$ or $[y, x]$
	to a set of intervals ${\cal I}$.
We constructed an overlap graph (circle graph) 
	from the set of intervals ${\cal I}$.
For example, from a sequence $(5,3,1,4,6,2)$,
	we construct a set of intervals 
	${\cal I}=\{[3, 5], [1, 4], [2, 6]\}$. 
For each $n \in \{5, 10, 30, \ldots ,  700\}$, 
	we generated 100 circle graphs with $n$ vertices
	and solved the coloring problems
   using CPLEX.

The results are summarized in Table~\ref{result1}.
The CL, AS, Ours columns represent
	the classical formulation, 
	asymmetric representative formulation,
	and our formulation (CG), respectively.
The ``\# $\omega =\chi$'' and ``\# $\chi_f =\chi$''columns 
	list the numbers of instances (out of 100 generated instances)
	satisfying $\omega(G)=\chi(G)$ and $\chi_f (G)=\chi (G)$, 
	respectively.
The ``max $\chi - \chi_f$'' column lists the maximum values 
	of $\chi (G) -\chi_f (G)$ over the 100 generated instances.	
We omit the computational results of some cases, denoted by ``-,''
	because of time limitation.
Table~\ref{result1} shows that
	our formulation solves coloring problems efficiently 
	compared to other formulations.

\begin{center}
\begin{table}[htb]
	\caption{Computational results for 100 randomly generated instances.}	\label{result1}
	\centering{
	\begin{tabular}{|l||r|r|r|r|r|r|r|r|r|} \hline
	 \multicolumn{1}{|c||}{$|V|$} &\multicolumn{1}{|c|}{$|E|$} &
	 \multicolumn{3}{c|}{computation time [s]} & 
	 \multicolumn{1}{c|}{\#} & 
	 \multicolumn{1}{c|}{\#} & 
	 \multicolumn{1}{c|}{max.}\\ \cline{3-5}
	   & & CL & AS & Ours &$\omega = \chi$ & $\chi_f=\chi$&$\chi-\chi_f$\\ \hline
	   5&3.35&0.013&0.004&0.006&100&100&0.0\\ \hline
	   10&14.17&0.018&0.006&0.012&95&95&0.5\\ \hline
	   30&145.39&0.038&0.016&0.021&94&94&0.5\\ \hline
	   50&413.40&0.364&0.059&0.029&89&91&0.7\\ \hline
	   70&802.79&2.096&0.158&0.045&92&94&0.7\\ \hline
	   100&1633.43&28.889&0.715&0.087&89&91&0.7\\ \hline
	   150&3784.46&- &4.092&0.202&82&83&0.8\\ \hline
	   200&6681.80&- &22.357&0.384&77&87&0.7\\ \hline
	   250&10412.11&- &- &0.710&77&85&0.7\\ \hline
	   300&14918.57&- &- &1.220&75&82&0.7\\ \hline
	   400&26404.89&- &- &2.988&71&77&0.8\\ \hline
	   500&41084.46&- &- &6.360&66&77&0.8\\ \hline
	   600&60191.42&- &- &12.430&70&83&0.8\\ \hline
	   700&81421.61&- &- &25.163&58&75&0.8\\ \hline	   
	\end{tabular}
	}
  \end{table}
\end{center}

We also generated hard instances $G(m)$
	proposed by Kostochka~(Section~6  
	in~\cite{kostochka2004coloring})
	for each	$m\in \{2,3,\ldots , 15\}$.
These results are summarized in Table~\ref{result2}.
When we employed the classical formulation and/or 
	asymmetric representative formulation,
	the execution time exceeded 1800s
	even in the case of $G(4)$.

\begin{center}
\begin{table}[htb]
	\caption{Computational results for hard instances $G(m)$.}	\label{result2}
	\centering{
	\begin{tabular}{|l||r|r|r|r|r|r|r|r|r|r|} \hline
	 \multicolumn{1}{|c||}{$m$} &
	 \multicolumn{1}{|c|}{$|V|$} &
	 \multicolumn{1}{c|}{computation time [s]} & 
	 \multicolumn{1}{c|}{$\omega$} & 
	 \multicolumn{1}{c|}{$\chi_f$} &
	 \multicolumn{1}{c|}{$\chi$} 
	 \\	 
	   & &  Ours & &&\\ \hline
2	&	24	&	0.031	&	5	&	6.500 	&	7	\\ \hline 
3	&	62	&	0.031	&	8	&	10.833 	&	11	\\ \hline 
4	&	122	&	0.141	&	10	&	15.750 	&	16	\\ \hline 
5	&	205	&	0.359	&	13	&	21.000 	&	21	\\ \hline 
6	&	316	&	1.468	&	15	&	26.833 	&	27	\\ \hline 
7	&	453	&	4.282	&	18	&	32.857 	&	33	\\ \hline 
8	&	617	&	13.531	&	20	&	39.062 	&	40	\\ \hline 
9	&	812	&	59.656	&	22	&	45.611 	&	46	\\ \hline 
10	&	1039	&	155.328	&	25	&	52.450 	&	53	\\ \hline 
11	&	1294	&	767.766	&	28	&	59.318 	&	60	\\ \hline 
12	&	1584	&	3878.625	&	31	&	66.500 	&	67	\\ \hline 
13	&	1904	&	9235.875	&	33	&	73.731 	&	74	\\ \hline 
14	&	2258	&	33087.641	&	35	&	81.143 	&	82	\\ \hline 
15	&	2647	&	119019.062	&	38	&	88.733 	&	89	\\ \hline 
	\end{tabular}
	}
  \end{table}
\end{center}

When we employed our formulation (CG), 
	CPLEX found optimal solutions for all instances 
	reported in Tables~\ref{result1} and~\ref{result2}
	at the root node (without any branching process).
Tables~\ref{result1} and~\ref{result2} show that
	all the generated instances	satisfy 
	$\chi (G) -1 < \chi_f (G) \leq \chi (G)$.
	
Ageev~\cite{ageev1996} constructed 
	a triangle-free graph $G_{\mbox{\scriptsize A}}=(V, E)$ 
	with chromatic number equal to 5, 
	where  $|V|=220$ and $|E|=1395$.
We calculated
	the fractional chromatic number of the graph 
	and obtained that
	$(\omega (G_{\mbox{\scriptsize A}}), 
		\chi_f (G_{\mbox{\scriptsize A}}), 
		\chi (G_{\mbox{\scriptsize A}}))
		=(2,  3.623\cdots,   5)$.
In this case, 
	the computational time required to solve problem CG 
	was 72.672s and
	number of branching nodes generated by CPLEX was 10,906.

\section{Capacitated Stowage Stack Minimization} \label{section:StowageStack}
\newcommand{\HGT}{\mbox{\rm Hgt}}
\newcommand{\deltaIN}{\delta^{\mbox{\scriptsize I}}}
\newcommand{\deltaOUT}{\delta^{\mbox{\scriptsize O}}}

In this section, 
	we briefly discuss an extension of our formulation
	to a capacitated stowage stack minimization problem.
Let $G=(V, E)$ and ${\cal I}(G)$ be
	a given circle graph 
	and the corresponding interval representation, 
	respectively.
Throughout this section, 
	$H$ denotes a positive integer representing ``capacity.''
For any independent set $S$ of $G$,
	the {\em height} of $S$ is equal to 
	the size of a maximum antichain of $(V, \preceq)$ 
	contained in $S$. 
In this section, 
	$\Xi_H$ denotes the set of independent sets (of $G$) 
	whose heights are less than or equal to $H$.
We introduce a 0-1 matrix $F_H$ indexed by 
	$V \times \Xi_H$ whose columns are 
	 the incidence vectors 
   of corresponding independent sets in $\Xi_H$.
We consider the following  
	0-1 integer programming problem:  
\[  \mbox{P$_H$:} \; \min \{ {\bm 1}^{\top} {\bm q}   
      \mid F_H {\bm q} = {\bm 1},  {\bm q} \in \{0,1\}^{\Xi_H} \}, 
\]
	where the variable vector ${\bm q}$ is indexed by 
	$\Xi_H$.
The above problem is essentially equivalent to 
	a capacitated stowage stack minimization 
	problem with zero rehandle constraint~\cite{wang2014stowage}.

We say that a $c$-coloring of $G$ is {\em $H$-admissible}
	if each color class is an independent set in $\Xi_H$.
It is obvious that each feasible solution of P$_H$  corresponds to 
	an $H$-admissible coloring. 
In a similar manner to that in Section~\ref{section:ILP},
	we introduce a directed graph and associate
	an $H$-admissible coloring of $G$ with a directed tree.
	
\begin{figure}[h]
\centering
\includegraphics[width=12cm]{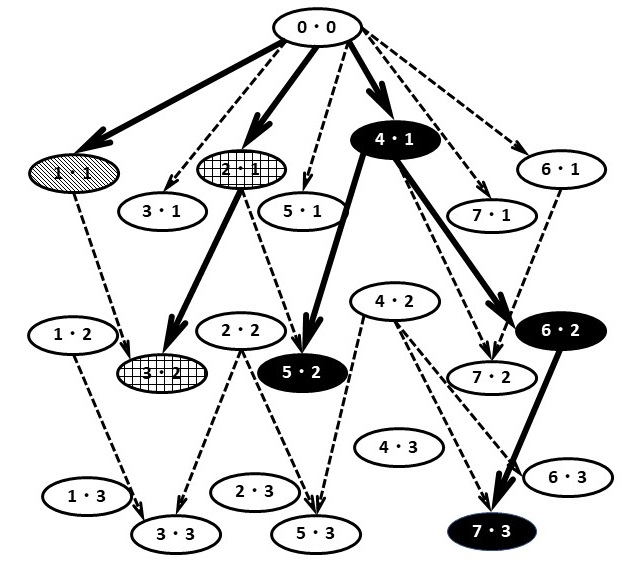}
\caption{Arc subset $T_3(\phi)$ in $\Gamma_3$ corresponding to 
	a 3-admissible 3-coloring $\phi$ in Figure~\ref{ex2}~(b).}\label{ex4}
\end{figure}

First, we define a directed graph $\Gamma_H=(V_H, A_H)$ as follows.
For each vertex $i \in V$, we construct a set of $H$ copies of $i$
	denoted by $V_H(i)=\{(i \cdot 1), (i \cdot 2), \ldots , (i \cdot H)\}$.
We introduce an artificial vertex $(0 \cdot 0)$ and define a vertex set 
	$V_H=\bigcup_{i \in V} V_H (i) \cup \{(0 \cdot 0)\}$.  	
The set of arcs $A_H$ is defined by 
\[
	A_H=\{((0 \cdot 0), (i \cdot 1)) \mid i \in V\}	
	\cup \{((i \cdot h), (j \cdot h+1))
		\mid I(i) \supsetneq I(j), h\in \{1,2,\ldots , H-1\} \}.
\]
From this definition, 
	it is clear that 
	$\Gamma_H=(V_H, A_H)$ is a directed acyclic graph.
Figure~\ref{ex4} shows $\Gamma_3$ corresponding to 
	the interval representation in Figure~\ref{ex1}~(c).

Given an $H$-admissible coloring $\phi'$, we define a subset of directed edges 
	of $A_H$ as follows.
Let $T(\phi' )$ be an arborescence of $\Gamma$, defined 
	in Section~\ref{section:ILP}.
For each vertex $i \in V$, 
	$\HGT (\phi', i)$ denotes the length 
	(number of edges) in a unique path in $T(\phi')$ 
	from the root $0$ to $i$. 
Obviously, we have  $\HGT (\phi', i) \leq H \; (\forall i \in V)$.
We define a set of arcs $T_H (\phi' )$ of $\Gamma_H$ by
\[
	T_H (\phi' )=\{(i \cdot h-1), (j \cdot h)
		 \mid (i, j) \in T(\phi' )\}
\]
where $h=\HGT (\phi' , j)$.
Figure~\ref{ex4} shows an arc set 
	$T_3(\phi)$ in $\Gamma_3$ corresponding to 
	the 3-admissible 3-coloring $\phi$ in Figure~\ref{ex2}~(b).

For any vertex $(i \cdot h) \in V_H$, 
	$\deltaIN (i \cdot h)$ and $\deltaOUT (i \cdot h)$
	denote a set of arcs in-coming to $(i \cdot h)$ in $A_H$ and 
	a set of arcs emanating from  $(i \cdot h)$ in $A_H$, respectively. 
Given an arc subset $T' \subseteq A_H$ 
	and vertex $(i \cdot h) \in V_H$, 
	we define a set of vertices
	$\CH (T', (i \cdot h))
		=\{j \in V \mid ((i \cdot h), (j \cdot h+1)) \in T'\}$.  
Then, we have the following property.

\begin{lemma} \label{arboriffH}
Let $T'$ be an arc subset of  $\Gamma_H$.
Then, there exists an $H$-admissible $c$-coloring $\phi'$ 
	of a given circle graph $G$ 
   satisfying $T'=T_H (\phi')$
	if and only if 
\begin{description}
\item[D0:] for any vertex $i \in V$, 
	$T'$ includes a unique arc in $ \bigcup_{h=1}^H \deltaIN (i \cdot h)$, 
\item[D1:] for each $(i \cdot h) \in V_H \setminus \{(0 \cdot 0)\}$,
 	$\CH (T', (i \cdot h))$ is a chain of $(V, \preceq)$ or the empty set;  
 	if  $\CH (T', (i \cdot h))$ is a (non-empty) chain, 
 	then $T'$ contains a unique in-coming arc to $(i \cdot h)$, and
\item[D2:] 
 the size of every antichain of $(V, \preceq)$ 
  contained in $\CH (T', (0 \cdot 0))$ is less than or equal to $c$.
\end{description}
\end{lemma}

\noindent
Proof (outline).
Given a set of arcs $T' \subseteq A_H$ 
	satisfying conditions D0, D1, and D2, 
	we define a set of arcs $T''$ of $\Gamma$ by
\[
	T''=\{(i, j) \in A \mid 
		\exists h \in \{0, 1, 2, \ldots ,H-1\},
		((i\cdot h), (j \cdot h+1)) \in T' \}.  
\]
Condition D0 implies that $T''$ is an arborescence of $\Gamma$.
From conditions D1 and D2, $T''$ satisfies conditions C1 and C2 
	in Lemma~\ref{arboriffH}, and thus there exists a $c$-coloring,
	denoted by $\phi''$, of $G$.
The definition of $\Gamma_H$ and condition D1 imply that 
	 $\HGT (\phi'', i) \leq H \; (\forall i \in V)$,
	 which implies that $\phi''$ is $H$-admissible.

The converse implication is obvious.
\QED

\medskip

Now, we give an integer linear programming formulation.
For each arc $((i \cdot h), (j \cdot h+1)) \in A_H$, 
	we introduce a 0-1 variable $x^{i\cdot h}_j$.
The vector of all 0-1 variables 
	is denoted by ${\bm x} \in \{0, 1\}^{A_H}$. 
For any vertex $(i \cdot h) \in V_H$,  
  $\RR{\bm x}{i \cdot h}$ denotes a subvector 
  of ${\bm x}$ indexed by 
  set of arcs $\deltaOUT (i \cdot h)$, 
  or   $\deltaOUT (i \cdot h) = \emptyset$.
Lemma~\ref{arboriffH} implies 
	a new formulation of P$_H$ as follows:

\begin{subequations}
	\label{CGH}
	\begin{alignat}{2}
		{\rm CG}_H\mbox{:} \: {\rm min.}\ & c \nonumber \\
		{\rm s.t.}\ & M\RR{\bm x}{0 \cdot 0} \leq \ c{\bm 1}, 
			&&  \nonumber \\ 
		& \RR{M}{i}\RR{\bm x}{i \cdot h} 
		\leq
		\left(
			\sum_{(j \cdot h-1) \in \deltaIN (i \cdot h)}x^{j \cdot h-1}_{i}
		\right)    {\bm 1} \quad
			&& (\forall (i \cdot h) \in V^{\bullet} \times \{1, \ldots , H-1\}),
			   \nonumber \\
		&  \sum_{h=1}^H \sum_{(i \cdot h-1) \in \deltaIN (j, h)}x^{i \cdot h-1}_j = 1\  
			&& (\forall j\in V),  \nonumber \\
		& x^{i \cdot h}_j\in \{0, 1\}\ 
			&& (\forall ((i \cdot h),(j \cdot h+1)) \in A_H), \nonumber \\
		& c\in \mathbb{Z}_+.  \nonumber 
	\end{alignat}
\end{subequations}

It is not difficult to show the following.
\begin{theorem}
The optimal value of the linear relaxation problem of CG$_H$
	is equal to the optimal value of the linear relaxation problem of P$_H$.
\end{theorem}
\noindent
Proof is omitted.

\section{Conclusion} \label{conclusion}
In this paper, 
	we proposed a new formulation for coloring circle graphs.
Our formulation is based on an interval representation of a given circle graph
	and uses a hierarchical structure 
	of a set of intervals corresponding to each independent set.
By employing  Dilworth's theorem,  we obtain a simple system 
	of inequality constraints represented by a clique matrix
	of an interval graph defined by a given interval representation. 
	
An advantage of our formulation is that
	the corresponding linear relaxation problem finds 
	the fractional chromatic number of a given circle graph.
Thus, 	our formulation also gives a polynomial-sized formulation 
	for a fractional coloring problem on a circle graph.
	
We confirmed by computational experiments that 
	a commercial IP solver can find a coloration quickly
	under our formulation.
When we employed our formulation, 
	CPLEX found optimal solutions for all the instances 
	randomly generated in our computational experiments
	at the root node (without any branching process).
The results of our computational experiments indicate that
	the chromatic number $\chi (G)$ of a circle graph $G$ is 
	very close to its fractional chromatic number $\chi_f (G)$.
We conjecture that
	there exists a constant $C$ satisfying 
	$\chi(G) -C \leq \chi_f (G)$ for any circle graph $G$.

We extended our result to 
	a formulation for a capacitated stowage stack minimization problem.
Future work is required to evaluate the computational performance
	of the proposed formulation.

\section*{References}
\renewcommand{\refname}{}

\end{document}